\documentclass[fleqn,usenatbib]{mnras}

\usepackage{newtxtext,newtxmath}

\usepackage[T1]{fontenc}

\DeclareRobustCommand{\VAN}[3]{#2}
\let\VANthebibliography\thebibliography
\def\thebibliography{\DeclareRobustCommand{\VAN}[3]{##3}\VANthebibliography}

\usepackage{graphicx}	
\usepackage{amsmath}	
\usepackage{soul}

\title[Constraining the BLR Size in NGC 1068]{Constrain the Size of Broad Line Region  in NGC 1068 Using Sub-millimeter  Excess}

\author[Qi Guo et al.]{Qi Guo,$^{1}$
Junzhi Wang,$^{1}$\thanks{E-mail: junzhiwang@gxu.edu.cn}
Lingrui Lin,$^{2,3,4}$
Yuqiang Li,$^{5}$
Zhi-yu Zhang$^{2,3}$
and Fengyao Zhu$^{6}$
\\
$^{1}$Guangxi Key Laboratory for Relativistic Astrophysics, School of Physical Science and Technology, Guangxi University, Nanning 530004,  PR China\\
$^{2}$School of Astronomy and Space Science, Nanjing University, Nanjing, 210093, PR China\\
$^{3}$Key Laboratory of Modern Astronomy and Astrophysics, Nanjing University, Ministry of Education\\
$^{4}$INAF - Arcetri Astrophysical Observatory, Large E. Fermi 5, 50125, Florence, Italy\\
$^{5}$Korea Astronomy and Space Science Institute, No. 776, Daedeok-daero, Yuseong-gu, Daejeon, Republic of Korea\\
$^{6}$Research Center for Computational Earth and Space Science, Zhejiang Laboratory, Hangzhou 311100, China\\
}
\date{Accepted 2026 October 12. Received 2026 October 10; in original form 2026 June 4}

\pubyear{\the\year{}}

\begin{document}
\label{firstpage}
\pagerange{\pageref{firstpage}--\pageref{lastpage}}
\maketitle

\begin{abstract}
Constraining the size of broad-line region (BLR) close to the accretion disk in active galactic nuclei (AGNs) is key to understand the central engine and  evolution of AGN. We propose a new method to constrain the BLR size,  based on millimeter to sub-millimeter spectral energy distribution (SED), from which a sub-millimeter excess may be found. Such excess can be used to determine the solid angle of the free-free emission region from BLR,  as a region with high electron density and optically thick at sub-millimeter wavelength for free-free continuum.
 Millimeter to sub-millimeter SED analysis toward  the central source S1 in a nearby type II AGN NGC 1068 was done, which gave an upper limit on the solid angle of BLR region as $(4.94 \pm 0.61)\times10^{-6}\ {\rm arcsec}^2$, corresponding to a radius no more than 0.085\,pc. Such a constraint on the BLR size is important for estimating the  masses of central supermassive black holes in AGNs.  
\end{abstract}




\section{Introduction}

Active galactic nuclei (AGNs), which are powered by supermassive black holes (SMBHs) accreting large amount of matter, release energy in the form of radiation across the electromagnetic spectrum from radio to X-ray wavelengths \citep{LR1971}. In the unified picture of AGNs, a central SMBH is surrounded by a sub-parsec-scale accretion disk,  broad-line region (BLR), narrow-line region (NLR), and an obscuring dusty torus \citep{Antonucci1993ARA&A}.

BLR is thought to be a dense ionized region located close to the central SMBH, producing broad emission lines with characteristic velocity widths of $\sim10^3$ to $\sim10^4\ \mathrm{km\ s^{-1}}$  \citep{Ho96ApJ}. The gas in  BLR typically has electron densities of $n_e \sim 10^{8}$-$10^{11}\,\rm{cm^{-3}}$  \citep{AO1988AJ,SLB2014MNRAS}. The size of BLR is hard to be determined. The BLR radius, defined as the distance between the accretion disk and the line-emitting gas, has been measured through reverberation mapping (RM) studies to range from several light-days to hundreds of light-days, corresponding to $\sim0.003$--$0.8\,\rm{pc}$  \citep{BM82ApJ}.  However, this method  is mostly limited to Type~I AGNs for probing BLR, as it relies on the direct observation of broad emission lines, mostly at optical wavelengths, while BLR can be detected with near-infrared observations in part of  Type~II AGNs without heavy obscuration \citep {Reunanen2002MNRAS}. 

Free-free emission should exist in BLR, as an ionized region.  The optical depth of  free-free emission,  $\tau_{ff}$, is about proportional to $\nu^{-2.1}$ \citep{Condon+92}. Therefore, BLR can be optically thick at millimeter (mm) to sub-millimeter(sub-mm) wavelengths, based on typical size and electron density of BLR. On the other hand, free-free emission from BLR should be optically thin at optical wavelength. Otherwise, broad H$\alpha$ line cannot be detected even in type I AGNs.  Thus, an excess close to black body at mm to (sub-)mm wavelength from the contribution of BLR can be expected for spectra energy distribution (SED) analysis. Millimeter to (sub-)mm wavelength excess had been found in many galaxies \citep{Condon+92,Garc+16}, and different origins have been proposed, including dust, free--free, and compact synchrotron emission\citep{Baskin21,Michiyama+23,Mutie2025}.
Even though it is still difficult to spatially resolve the BLR,  mm to sub--mm SED for the central source of the AGNs with high-resolution measurements, can help to rule out extended warm or cold dust emission. Although contributions from unresolved hot dust and other compact nuclear components cannot be completely excluded, an upper limit on sub-mm thermal emission from BLR can be obtained. 

NGC~1068, a nearby bright prototypical Seyfert~II galaxy with sub-mm excess at a redshift of $z = 0.00348$, located at a distance of $D_{L} = 13.97 \pm 2.10 \, \rm{Mpc}$ ($1'' \approx 68 \pm 10 \,\rm{pc}$)  \citep{Anand+21}, is a good choice for such SED analysis.  
In this Letter, we present a novel method to constrain the physical size of BLR in NGC 1068, utilizing the sub-mm excess detected in Atacama Large Millimeter/submillimeter Array (ALMA) observations spanning Bands 3 to 9.

\section{ALMA OBSERVATIONS AND DATA REDUCTION}

\begin{table*}

\centering
\footnotesize
\setlength{\tabcolsep}{3pt} 
    \begin{center}
\caption{Summary of ALMA data used in this work}
\label{table1}
\begin{tabular}{cccccccccc}
\hline
\hline
Band & Frequency & Project ID &Observation Date& On source time & Flux &Flux Calibrator& Calibration error$^{a}$&Beam size&PA \\
&(GHz)&&&(min)&(mJy)&&(\%)&($''$)&($^\circ$)\\
\hline
3 & 84.7 & 2018.1.01135.S & 2019 June 22&48.12&15.1$\pm$ 0.5& J0238+1636&3.3 & 0.06$\times$ 0.05&-38.1\\
3 & 98.2 & 2018.1.01135.S & 2019 June 22&48.12&13.8$\pm$ 0.4& J0238+1636&3.3 & 0.05$\times$ 0.04&-38.1\\
6 & 266.2 & 2016.1.00052.S & 2017 Sept 11& 37.73&12.4$\pm$ 0.5&J0006-0623& 3.6&0.05$\times$0.04&68.8\\
7 & 322.7 & 2019.1.00854.S & 2021 Sept 27 & 27.20 &12.0 $\pm$ 0.5&J0238+1636&  3.8&0.03$\times$0.03&70.9\\
7& 356.3 &2019.1.01540.S&2021 Sept 4& 82.78&12.9$\pm$0.3 & J0423-0120& 2.2&0.02$\times$0.02&39.8\\
8 & 477.4 &2021.1.00279.S & 2021 Nov 20& 34.97&16.1 $\pm$ 0.6&J0423-0120&2.9&0.05$\times$0.04&84.5\\
9 &684.9& 2022.1.00017.S &2023 June 2& 31.95&21.6 $\pm$ 2.2&J2253+1608&4.7&0.04$\times$0.04&-74.5\\
\hline
\end{tabular}
\begin{flushleft}
\footnotesize
$^{a}$ Calibration uncertainties are derived from the flux calibrator monitoring measurements; see Section~2 for details.
\end{flushleft}
    \end{center}
\end{table*}

To ensure that the SED is mainly from the central region excluding cold dust contribution as much as possible, we selected archival ALMA datasets with angular resolutions better than 0.05\arcsec. The archival data were obtained from multiple ALMA projects (see Table \ref{table1} for details). Representative maps for Bands 3, 6, 7, 8 and 9 were constructed by combining these high-resolution archival data with our newly obtained 3 mm observations (Project ID: 2018.1.01135.S, PI: Junzhi Wang).

The data were manually calibrated using CASA (version 6.5.4). Since flux calibrators exhibit intrinsic variability, their flux densities and associated calibration uncertainties were determined using the ALMA Calibrator Source Catalogue, which provides regularly updated monitoring measurements. We adopted calibrator measurements obtained closest in time to the observation date and restricted our sample to datasets for which suitable calibrator monitoring measurements were available, ensuring reliable flux estimates. For Bands 3, 6, 7, and 9, calibrator flux densities and their uncertainties were adopted from monitoring measurements close in frequency and time to the observations, and the corresponding spectral indices were derived from these measurements. For Band 8, no suitable same-band monitoring measurement was available. We therefore constructed calibrator light curves at two lower frequencies and interpolated them to the date of observation. The interpolated flux densities were then used to derive a spectral index, from which the Band 8 calibrator flux density and its uncertainty were estimated.

The calibrated uv data were imaged using the \texttt{tclean} task in CASA. The data were first imaged in cube mode to identify spectral line emission within the observed frequency range. Although the strong molecular line emission mainly arises from the CND rather than the S1 region, channels contaminated by line emission were flagged and excluded prior to continuum imaging to obtain a cleaner continuum dataset.
In particular, channels contaminated by the HCN $J=3$--$2$ and HCO$^+$ $J=3$--$2$ emissions from the CND in project 2016.1.00052.S \citep{Imanishi2018ApJ}, as well as by $^{13}$CO $J=3$ -- $2$, HCO$^+$ $J=4$--$3$ and [C \,{\sc i}]  (${}^3P_1$--${}^3P_0$) in other datasets, were were excluded.
 The remaining line-free channels were used to produce continuum images via multi-frequency synthesis (mfs). During the imaging process, primary-beam-based masking with a threshold of pb$>0.2$  was applied and a Briggs weighting scheme with a robust parameter of 0.5 was adopted.

The source position was determined by performing a Gaussian fit to the continuum images using the CASA task {\tt imfit}. Flux densities were measured using aperture photometry with a common circular aperture radius of 0.07\arcsec\ across all bands to ensure consistency in the derived SED.
The associated uncertainties include  contributions from both flux calibration, estimated from calibrator monitoring, and measurement errors.

\section{result} 

\begin{figure}
\centering
\includegraphics[width=0.45\textwidth]{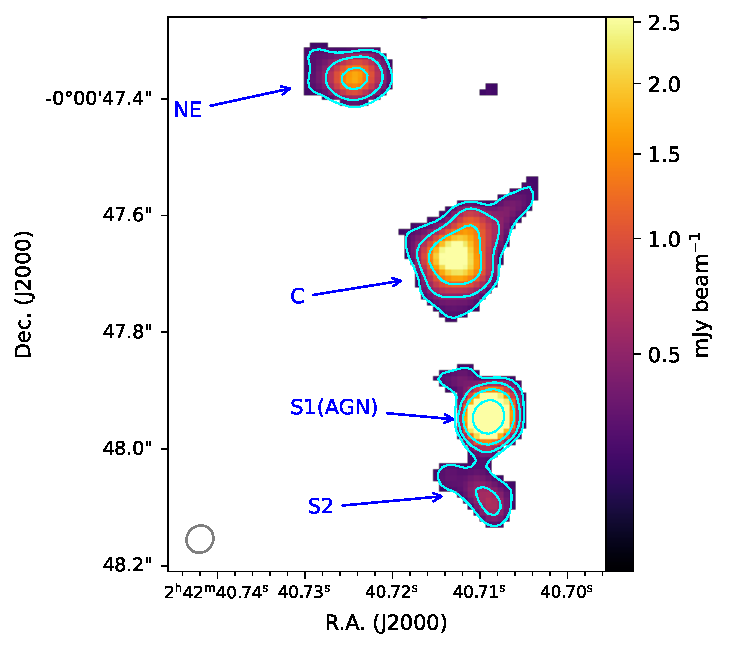}
\caption{ALMA 3\,mm continuum image of NGC~1068 with an angular resolution of $0.05''$. Contours are plotted at [10, 20, 50, 180]$\sigma$ levels. The labeled components correspond to the NE source, C component, S1 (AGN), and S2 source. synthesized beam is shown in the lower-left corner.}
\label{fig:3mm}
\end{figure}

The physical resolution of our 3\,mm observations is $0.05'' \approx 3.4$ pc. Four distinct radio components are identified: NE, C, S1, and S2. They are consistent with those reported in the literature \citep{GSC04} , as shown in Figure ~\ref{fig:3mm}. Among them, NE, C, and S2 are associated to the radio jet and are primarily contributed by synchrotron emission, while the AGN is identified with the compact component S1  \citep{GSC04}, as also shown in Figure~\ref{fig:3mm}.

SED  analysis  from mm to sub-mm wavelength  is only done for  the nuclear component S1. Although the SED of S1 in NGC 1068 has been extensively studied across multiple wavelengths (e.g.,  \citealt{HPB2008}, \citealt{Mutie2025}),its mm to sub-mm SED remains poorly constrained, with multiple emission mechanisms still under debate (e.g.,  \citealt{Garc+16}, \citealt{Baskin21}, \citealt{IKD2020ApJ}). This can also be caused by the large uncertainties of absolute flux calibration, because of the lack of monitoring toward flux calibrators  for most observations used in the literature. 

\begin{figure}
    \begin{centering} 
\includegraphics[width=0.46\textwidth]{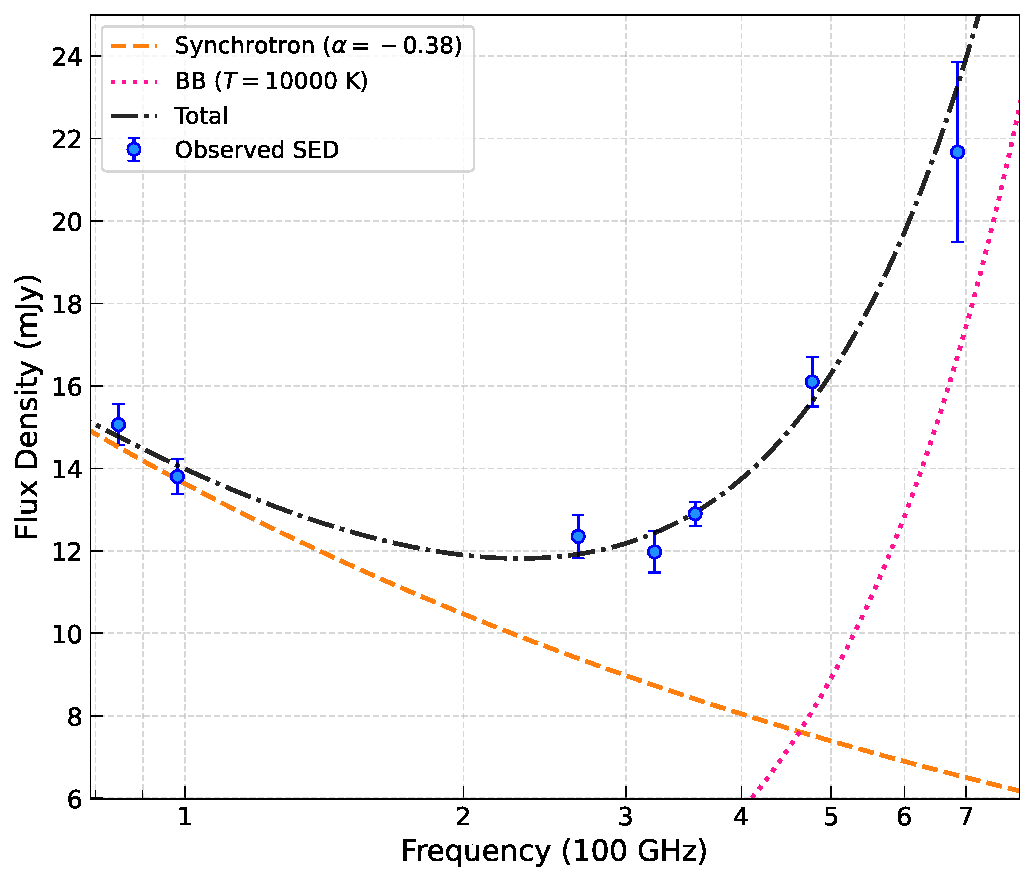}
\caption{The SED of the AGN (S1) in NGC 1068.
The flux densities were derived using aperture photometry with the same aperture size for all measurements. The errors are estimated by 
combining calibration and measurement errors from the Gaussian fit. The black dash--dotted line represents the two-component model, consisting of synchrotron emission and optically thick free--free emission modeled as a blackbody spectrum.
The orange dashed line is the synchrotron power low emission with $\alpha=-0.38\pm 0.06$. 
The pink dotted line indicates a blackbody with $ T = 10^4 \,\rm K$.
}
\label{fig:2}
\end{centering} 
\end{figure}

SED over the frequency range of $84$ -- $685\,\rm{GHz}$  was obtained. For each data point, a continuum image was independently produced, achieving a synthesized beam size of $0.02''$-- $0.06''$, corresponding to $1.36$--$4.08\,\rm{pc}$ in  physical scale. 
Flux densities were measured from the images,  while  the associated uncertainties include contributions from both the uncertainties of  flux calibrators  and measurement error (see Table~\ref{table1}). The SED result exhibits an initial power-law decline, followed by a
spectral upturn at higher frequencies (see Figure ~\ref{fig:2}).

The SED can  be well described by a two-component model consisting of synchrotron and sub-mm excess as blackbody emission  (see Figure ~\ref{fig:2}), with a reduced chi-square value of
$\chi_\nu^2 = 0.85 $. 
The low-frequency spectrum is fitted by a synchrotron power-law, $S_\nu \propto \nu^\alpha$, with a spectral index of $\alpha = -0.38\pm 0.06$, while the high-frequency component can be described  with optically thick free--free emission as a black body spectrum $S_{\nu}=\Omega B_{\nu}(\rm T)$, assuming a a temperature of $T = 10^4\,\rm{K}$, $\Omega = (4.94\pm 0.61)\times10^{-6}\ {\rm arcsec}^2$.

\section{discussion}

The SED indicates that synchrotron emission with a relatively flat spectrum  dominates  the emission at 84--300\,GHz. Such a flat synchrotron spectrum ($\alpha > -0.5$) is expected in compact AGN core regions due to synchrotron self-absorption and the superposition of multiple emitting zones  \citep{RL16}.

Assuming a  spherical symmetry or spherical shell, the derived emitting area corresponds to a BLR radius of  ($R_{\rm BLR}$)  $\sim 0.085$ pc, estimated from $R_{\rm BLR}=D\sqrt{\Omega/\pi}$ at the distance ($D$) of $\sim$14 Mpc \citep{Anand+21}. 
Since this sub-mm excess can also be from hot dust, the estimated radius of BLR can only be considered as the upper limit.  Based on the AGN luminosity of NGC 1068 $L_{\rm bol} \sim 10^{45} \, \rm{erg\, s^{-1}}$ \citep{Woo+2002ApJ}, and the empirical BLR radius--luminosity relation\citep{Bentz2013ApJ}, a characteristic BLR size in NGC~1068 is estimated to be $ R\sim 0.03$-$0.05\, \rm{pc}$, which is consistent with the value derived  from sub-mm excess.

We now return to the question of whether the free-free emission from BLR of NGC 1068 is optically thick  at sub-mm wavelength. Free--free absorption coefficient at radio wavelength can be used that  from \cite{Gordon2002},
\begin{equation}
\kappa_C = 9.770\times10^{-3}\frac{n_en_i}{\nu^2T_e^{3/2}}\left[17.72+\ln\left(\frac{T_e^{3/2}}{\nu}\right)\right]
\end{equation}
where $n_e$ and $n_i$ are electron and positive ions densities in units of cm$^{-3}$, $\nu$ is frequency in Hz, and T$_e$ in K, while 17.72+ln($\frac{T^{3/2}_e}{\nu}$) is from numerical result of Gaunt factor.  Together with $\tau_{ff}=\kappa_C L$, and reasonable assumptions of $T \sim 10^4$ K, $n_i\simeq$ $n_e$ for first order approximation and $n_e \sim 10^8\ \rm{cm^{-3}}$,  $\tau_{ff}$  is about 420  at 684.9 GHz (or 444$\mu$m) for the radius of 0.085 pc. Since  $\tau_{ff}$  is proportional to $n_e^2$ \citep{Condon+92,Gordon2002}, a higher electron density ($n_e>10^8 \,\rm cm^{-3}$) would result in an even larger optical depth. Furthermore, if the BLR is clumpy, $\tau_{\rm ff}$ would still exceed unity provided that the characteristic size of an individual clump is larger than 1\% of the total emitting region. Since the free--free emission from the BLR is optically thick, broad H26$\alpha$ emission is not expected under LTE conditions, consistent with its non-detection \citep{Izumi+16MN}.

On the other hand, since  $\tau_{ff}$ is proportional to $\nu^{-3}$ at near IR and optical wavelengths, the Gaunt factor is $\simeq$1 \citep{OF+2006agna.book} and (1-e$^{-h\nu/kT}$) is close to 1 instead of $\simeq$$h\nu/kT$ at radio wavelengths,  $\tau_{\rm ff}$ is about 0.01 at 4$\mu$m, or less than 1$\times10^{-4}$ at 6563~\AA\  for H$\alpha$, which means free-free absorption is negligible for H$\alpha$ emission in the BLR.

Free-free emission and absorption from several GHz to THz in NGC 1068 and other AGNs had been predicted by \cite{Baskin21}.  Free-free absorption is important below 10GHz for NLR and below 100GHz for BLR, while free-free emission will be important above 300 GHz in S1 from BLR of NGC 1068 \citep{Baskin21}, which is consistent with our prediction here. The major difference between our prediction to  that of \cite{Baskin21} is the electron density of BLR, where we used the typical value as 10$^{8}$cm$^{-3}$, while about varied values with distance to the central SMBH close to 10$^{7}$cm$^{-3}$ was used in \cite{Baskin21}. Never the less, it is still optically thick at $\sim$650 GHz if $n_e$ is 10$^{7}$cm$^{-3}$, which means the estimated $\Omega$ and radius of BLR  are correct.


Broad-band SEDs of S1 from centimeter to sub-millimeter have also been constructed by \cite{Michiyama+23} and \cite{Mutie2025}. The former study lacked data between 400 and 600 GHz. In both studies, most of the flux calibrators for the ALMA observations above 400 GHz were not monitored for flux variability. This is important because the S1 flux densities at these frequencies determine whether the SED shows a flattening from $\sim500$ to $\sim706$ GHz \citep{Mutie2025}. Thus, whether the reported flattening or decline over this frequency range is real remains uncertain, and an additional continuum emission component invoked to explain it may not be necessary. Nevertheless, a contribution from other components, such as coronal synchrotron emission, cannot be ruled out.

Cold and warm/hot  dust may also contribute to the sub-mm SED. However, high-resolution ALMA observations indicate that the dust contribution from S1 is relatively small, while much of the dusty molecular structure is distributed on larger spatial scales \citep{GB2019A&A}. Our $\sim0.05\arcsec$ observations further reduce contamination from such extended dust emission. In contrast, a contribution from unresolved hot dust close to the nucleus is more difficult to exclude.
As pointed out before hot dust  with temperature at the level of several hundreds K, can show similar feature in the  SED at sub-mm wavelength, since the peak of the a 300-400 K blackbody is at near-IR wavelengths. However, the emissivity of dust is normally  proportional to $\nu^{\beta}$, where $\beta$ is dust opacity spectral index between  1 to 2 \citep{Draine2006ApJ}. Dust  opacity at mm to sub-mm wavelengths is normally less than unity, which means the dust  continuum emission $S_\nu$ will be steeper than 2 as the blackbody spectrum, while the SED in  S1 of NGC 1068 (see Figure 2) presents a spectrum close to 2 within uncertainty.   The hot dust in S1 close to the center of in NGC 1068  was thought to have a temperature of $\sim320$ K \citep{Jaffe2004}, which means the brightness temperature of dust emission at sub-mm wavelength should be no more than 300 K,  or only be at the level of tens of K, depending on  dust opacity.  The measured sub-mm excess of $\sim$ 16 mJy at 684.9 GHz corresponds to 14.7 K,  at the spatial resolution of $0.04''$. So it is hard to exclude the possibility of hot dust emission with current data in this way, while observations at higher spatial resolution to $\sim 0.008''$ with ALMA can do. 

Any contribution from unresolved hot dust or compact coronal synchrotron emission would reduce the fraction of the observed sub-mm excess attributable to BLR free--free emission and hence lead to a smaller inferred BLR size. We therefore conservatively interpret the inferred radius of $\sim0.085$ pc as an upper limit on the BLR size.

High spatial resolution multi-wavelength observations with ALMA toward other nearby  AGNs, which can provide SED analysis  from mm to sub-mm for central region near SMBHs, will be an unique tool for placing an upper limit on the size of BLR in type I and II AGNs.  Although the flux calibration for high-frequency high-resolution observations may not be perfect, the results will not affect the main results of estimating size of BLR region, because the size of BLR  is proportional to square root of the flux. 

With size of BLR estimated from sub-mm excess, and line width obtained with H$\alpha$ \citep{Peterson2004}, broad Br$\gamma$ line \citep{Reunanen2002MNRAS}, or sub-mm radio recombination lines \citep{Wang2023MNRAS}, the mass of SMBH can be estimated with the  virial theorem. Thus, the sub-mm excess measured with mm to sub-mm wavelength SED provides  an important measurement for deriving mass of SMBH  in AGNs. Future high-resolution SKA observations may provide complementary constraints on the nuclear emission components at radio wavelengths for other AGNs, especially for high-$z$ sources to measure  emission for rest frame at millimeter \citep{KS2026}.

\section{Summary and Conclusion remarks}

 With typical size and electron density of BLR in AGNs, the free-free emission should be optically thick at sub-mm wavelengths, which can produce an excess in SED.  Such sub-mm wavelength excess can be used to constrain an upper limit on the size of BLR. An SED analysis from mm to sub-mm wavelengths toward the central source of NGC 1068, a nearby Type II AGN, was performed, yielding a constraint on the BLR radius of no more than 0.085 pc. This method can be used to constrain BLR sizes in local AGNs using high-resolution millimeter-to-submillimeter observations.

\section*{Acknowledgements}

This work is supported by National Key R$\&$D Program of China under grant 2023YFA1608204, the National Natural Science Foundation of China grant 12173067 and the Guangxi Talent Programme (Highland of Innovation Talents).  This letter makes use of the following ALMA data: ADS/JAO.ALMA\#2016.1.00052.S, 2018.1.01135.S, 2019.1.01540.S, 2019.1.00854.S, 2021.1.00279.S, 2022.1.00017.S. ALMA is a partnership of ESO (representing its member states), NSF (USA), and NINS (Japan), together with NRC (Canada), NSTC and ASIAA (Taiwan), and KASI (Republic of Korea), in cooperation with the Republic of Chile.

\section*{Data Availability}

The data underlying this article are available in the ALMA Science Archive. The observations analyzed in this work were obtained from archival ALMA projects. The calibrated datasets and derived continuum images used in the analysis are available from the corresponding author upon reasonable request.



\bibliographystyle{mnras}
\bibliography{BLR} 




\bsp	
\label{lastpage}
\end{document}